\documentclass[a4paper]{jpconf}
\usepackage{graphicx}
\usepackage[usenames]{color}
\begin{document}
\title{Bottom friction models for shallow water equations: Manning's roughness coefficient and small-scale bottom heterogeneity\footnote{IOP Conf. Series: Journal of Physics: Conf. Series 973 (2018) 012032. Doi: 10.1088/1742-6596/973/1/012032}}

\author{Dyakonova T.A., Khoperskov A.V.}

\address{Volgograd State University, Department of Information Systems and Computer Simulation, Volgograd, 400062, Russia}

\ead{khoperskov@volsu.ru}

\begin{abstract} %до 150 слов
 The correct description of the surface water dynamics in the model of shallow water requires accounting for friction. To simulate a channel flow in the Chezy model the constant Manning roughness coefficient is frequently used. The Manning coefficient $n_M$ is an integral parameter which accounts for a large number of physical factors determining the flow braking. We used computational simulations in a shallow water model to determine the relationship between the Manning coefficient and the parameters of small-scale perturbations of a bottom in a long channel. Comparing the transverse water velocity profiles in the channel obtained in the models with a perturbed bottom without bottom friction and with bottom friction on a smooth bottom, we constructed the dependence of $n_M$ on the amplitude and spatial scale of perturbation of the bottom relief.
\end{abstract}

\section{Introduction}
The models of shallow water are successfully applied for calculation of a wide range of geophysical currents including the dynamics of surface and groundwater, landslides \cite{shokina2015numerical}, torrential flows \cite{agafonnikova2017computer, shokin2016combined}, the pyroplastic and granular masses movement \cite{Cui2013}, the simulation of large-scale atmospheric, sea and ocean currents \cite{Majda2003}, and the technospheric security problems \cite{Luo2017, kulagin2016physical}. To simulate the dynamics of channel and floodplain currents in the shallow water models, it is essential accounting for the hydraulic resistance \cite{Baryshnikov2010,Dyakonova2014,Pisarev2013,Jong-Seok2017,habibi2014experimental}. The latter can be described using a number of developed phenomenological models \cite{habibi2014experimental}. The Chezy model, proposed as far back as the 18th century, is the most common \cite{Jong-Seok2017}. There are several expressions determining the Chezy coefficient, but the Manning formula containing the roughness coefficient $n_M$ \cite{Baryshnikov2010} is the most often used in practice. The parameter $n_M$ accounts for the features of both small-scale low heterogeneity and watercourse morphology on a large scale. The Manning roughness coefficient is a composite value generally taking into account the influence of vegetation, various obstacles, meandering, turbulence and other factors \cite{Doncker2009,Hadiani2013,Herget2010,Mohammadi2014}. Since the coefficient $n_M$ widely varies for different watercourses \cite{Pisarev2013} choosing its value in practice usually causes difficulties. In the Refs.~\cite{Hatzigiannakis2016,Hatzigiannakis2014,Lyra2010} the estimates for $n_M$ obtained from the Manning equation are given for various rivers. The coefficient $n_M$ is assumed to be a function of coordinates and may depend on the water layer thickness, $H$, under the conditions of the nonstationary flow \cite{khrapov2013mech}.

 It should be noted, that in the engineering applications of the shallow water model the Manning coefficient (or similar parametr) should be determined as correctly as possible for the existing small-scale inhomogeneity of the bottom relief \cite{agafonnikova2017computer,khrapov2013mech,nistoran2017one,Voronin2017}.
The main aim of this article is the research of the Manning coefficient $n_M$ dependence on the characteristic amplitude and spatial scales of the irregular structure of the channels relief. The estimations of the bottom friction coefficient for the Chezy formula have been obtained in a large series of the computer simulations results within the shallow water model. In particular, the relationship between the parameters of the small-scale bottom inhomogeneity in the channel and the Manning coefficient has been revealed.

\begin{figure}
\begin{center}
	\includegraphics[width=0.55\textwidth]{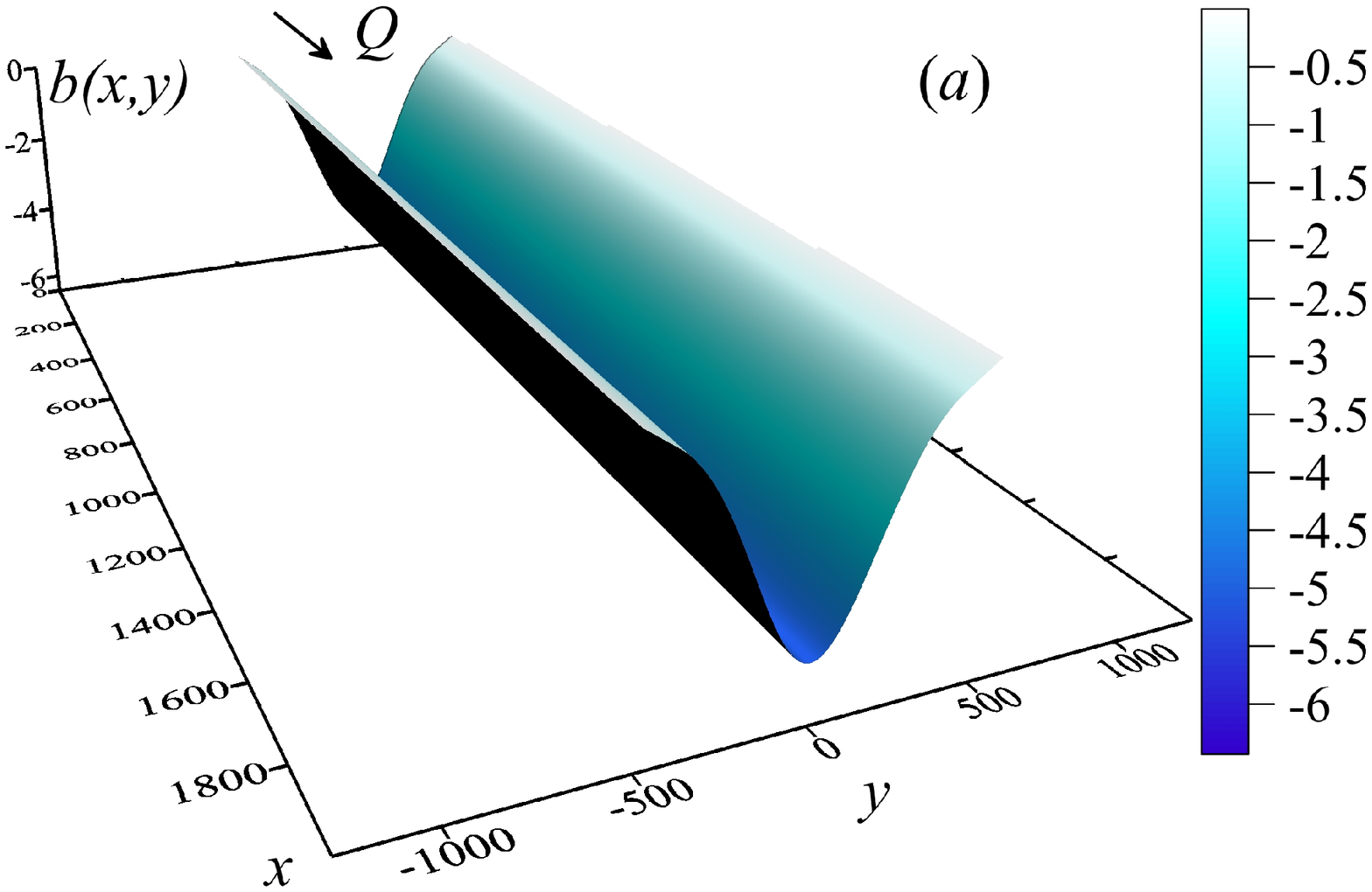}
    \includegraphics[width=0.54\textwidth]{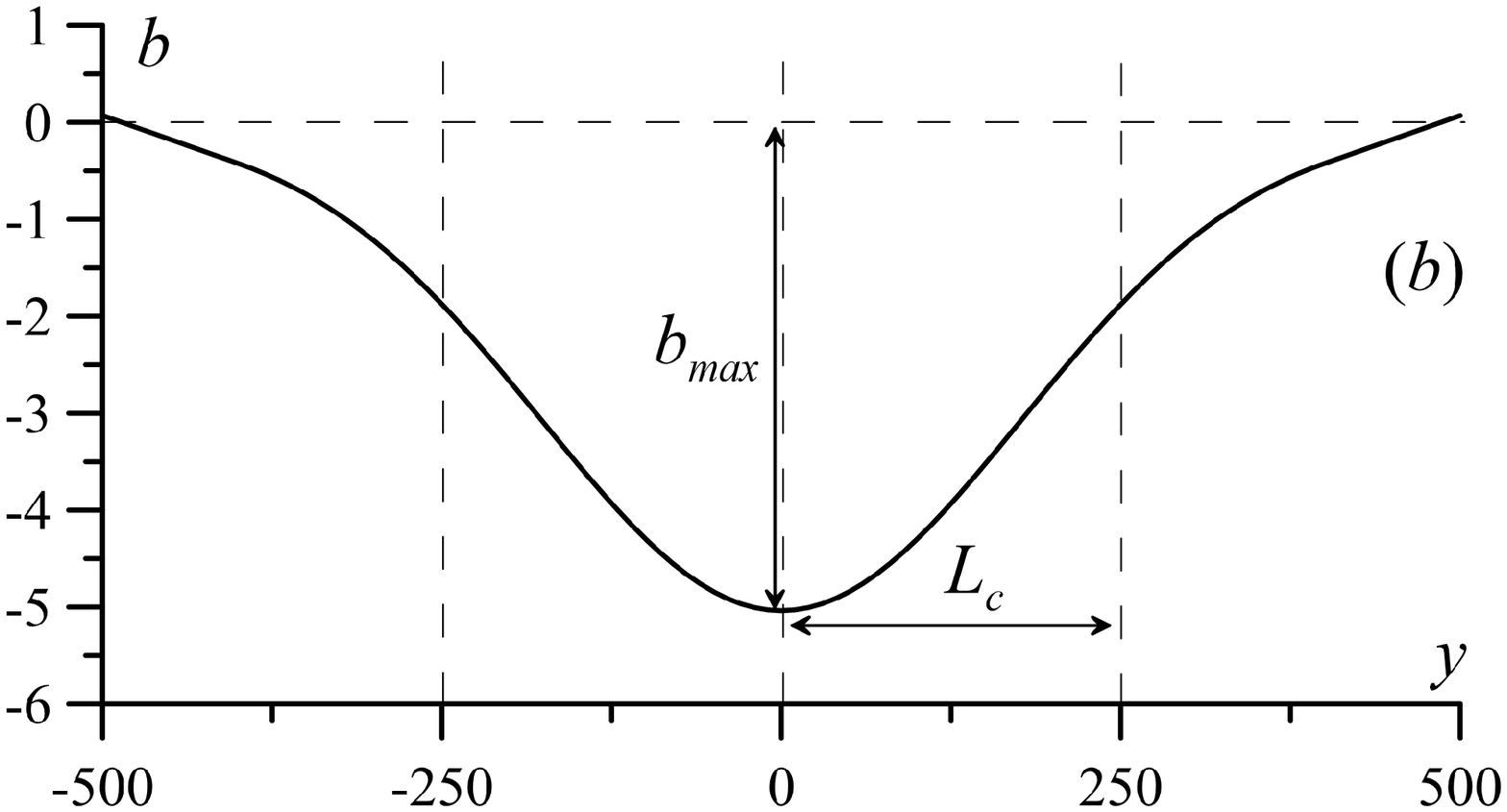}
\end{center}
\caption{\label{FIG01Dyakonova}The geometry of the computed area with a watercourse (meters are used as the coordinate units). The fixed volume of water per time unit $Q$ ($m^3/$~sec) is transferred to the watercourse input.}
\end{figure}

\section{Materials and Methods}

\subsection{Mathematical model}

The hydrodynamic model of the surface water dynamics should account for the heterogeneity of the relief, external and internal forces acting on the water layer, evaporation from the water surface, and filtration into the soil \cite{Dyakonova2014}. Our model utilizes the following Saint-Venant equations:
\begin{equation}\label{eq-SV-H}
\frac{\partial H}{\partial t}+\frac{\partial HV_x}{\partial x}+\frac{\partial HV_y}{\partial y}=\sigma,
\end{equation}
\begin{equation}\label{eq-SV-ux}
\frac{\partial V_x}{\partial t}+V_x\frac{\partial V_x}{\partial x}+V_y\frac{\partial V_x}{\partial y}=-g\frac{\partial \eta}{\partial x}+F_x+\frac{\sigma}{H} (U_x-V_x),
\end{equation}
\begin{equation}\label{eq-SV-uy}
\frac{\partial V_y}{\partial t}+V_x\frac{\partial V_y}{\partial x}+V_y\frac{\partial V_y}{\partial y}=-g\frac{\partial \eta}{\partial y}+F_y+\frac{\sigma}{H} (U_y-V_y),
\end{equation}
where $H$ is the water depth, $V_{x}, V_{y}$ are the components of the water velocity vector $\vec{V}$ averaged over the vertical coordinate, $\sigma$ is the surface density of water sources and sinks (m/sec), $g$~is the gravitational acceleration, $\eta(x, y, t) = H (x, y, t) + b (x, y)$ is the vertical coordinate of the water surface, $b(x, y) $ is the relief function, $U_{x}, U_{y} $ are the components of the water velocity vector in the source or drain $\vec{U}$, $F_{x}, F_{y}$ are the components of the force vector $\vec{F}$.

Neglecting the internal viscosity and interaction with the surface layer of the atmosphere the expression for the force can be written as:
\begin{equation}
\vec{F}=\vec{F}^{fric}+\vec{F}^{cor},
\end{equation}
where $\displaystyle \vec{F}^{fric} = -\frac{\lambda}{2}\vec{V}|\vec{V}|$ is the bottom friction force, $\displaystyle \lambda = \frac{2gn^{2}_M}{H^{4/3}}$ is the coefficient of hydraulic resistance, $n_{M}$ is the Manning's roughness coefficient (sec/m$^{1/3}$), $\vec{F}^{cor} = 2 [\vec{V} \times \vec{\Omega}]$ is the Coriolis force, $\vec{\Omega}$ is the angular velocity of the Earth's rotation.

\subsection{Numerical model}
The numeric solution of the Saint-Venant equations has been obtained using the CSPH-TVD (Combined Smooth Particle Hydrodynamics -- Total Variation Diminishing) scheme, which has the second order of accuracy in time and space, conservativeness, and well balance. Its advantage is the possibility of the through calculations of the unsteady ``water -- dry bottom'' boundaries on a substantially inhomogeneous relief.
For the boundary regions where the water leaves the calculation zone ($x = x_{\max}$, See Figure 1) we have specified the boundary conditions by the method described in Ref.~\cite{Dyakonova2016boundary}.
The essence of the CSPH-TVD method has been explained in detail in Refs.~\cite{khrapov2013mech,Khrapov2011}. The calculation is carried out on a regular grid with cells $\Delta {x} = \Delta {y}$. The ``liquid'' particles are placed inside the cells. In accordance with the Lagrangian approach, the changes in the characteristics of the particles are calculated under the action of hydrodynamic and internal forces. At the Euler stage, the characteristics with respect to the flows through the cell's boundaries are refined. Since this algorithm is suitable for parallelization, we utilized the NVIDIA CUDA GPU technology in the developed software~\cite{Dyakonova2016}.

Figure 1 shows a watercourse situated along the $x$-coordinate which is specified by a function of bottom $b(x,y)$. The figure \ref{FIG01Dyakonova}\,\textit{a} demonstrates only a part of the watercourse, because its total length amounts 20480~m according to our model.
The shape of the watercourse's bottom is chosen to be an exponential function:
\begin{equation}\label{eq_b0_Dyakonova}
b_0(x,y) = -b_{\max}\cdot \exp(-y^2/L_c^2)\cdot (1+Ix)
\end{equation}
where $b_{max} = 5$\,m, and $L_c = 250$\,m. A small modification of Eq.~(\ref{eq_b0_Dyakonova}) for $|y|> L_c$ provides a slow linear growth of $b_0(y)$ outside of the main part of the watercourse.
The coefficient $I$ (m/km) in Eq.~(\ref{eq_b0_Dyakonova}) provides a weak slope of the watercourse's bottom along the $x$-coordinate.
We name function $b_0(x, y)$ the smooth bottom profile, and in all our calculations it is assumed to be unchanged. We have chosen typical for the large plain rivers value of slope $I = 0.07$\,m/km (for example,  for the Volga, Don, Dnepr, Oka, Amazon, and Madeira the slope lies within $0.01 \leq I \leq 0.1$, for the Western Dvina, Neman, Kama, Seversky Donets, Mississippi, and Lower Vistula it is within $0.1 <I \leq 0.2$, and for the Thames and Congo it is within $0.2 <I \leq 0.4$).

The length of the numerical cell is $\Delta {x} = \Delta {y} = 10$\,m, thus for the computational domain with number of cells $2048 \times 256$  the total size of the computing area is 20480~m $\times$~2560~m.

A fixed source of water is maintained at the entrance of the watercourse with the capacity $Q = 1500$\,m$^3$/sec, which roughly corresponds to the water discharge values for large plain rivers (for example, the Don River $Q = 680$\,m$^3$/sec, Oka River $Q = 1260$\,m$^3$/sec, Kama River $Q = 4000$\,m$^3$/sec, Rhine River $Q = 2200$\,m$^3$/sec, Vistula Rive $Q = 1080$\,m$^3$/sec). We have set the water volume $\Delta {V^{n}} = Q \Delta{t^{n}}$ for the area $S_Q = 50$~m$^2$ at the left boundary inside the watercourse ($x=0$, See Figure~1\,a) at each time step. The particular choice of $S_Q$ does not influence on final results.
The value $Q = 1500$~m$^3$/sec ensures the water depth in the fairway ($y = 0$) within $H_{\max} \sim 2 \div 4$~m depending on other factors.
The calculation is conducted up to $t_{\max} = 144000$~sec, which certainly guarantees the establishment of a quasi-stationary solution.

Water does not reach boundaries $y = y_{\max}$ and $y = y_{\min}$ in our simulations, thus we have set the boundary conditions of the ``water-dry bottom'' type within the calculation area. At the boundary $ x = 0 $ we have put a solid wall and set the conditions of the waterfall type to $ x = x_{\max} $ (see the details in~\cite{Dyakonova2016boundary}).

\begin{figure}[th!]
	\begin{center}
		\includegraphics[width=0.98\textwidth]{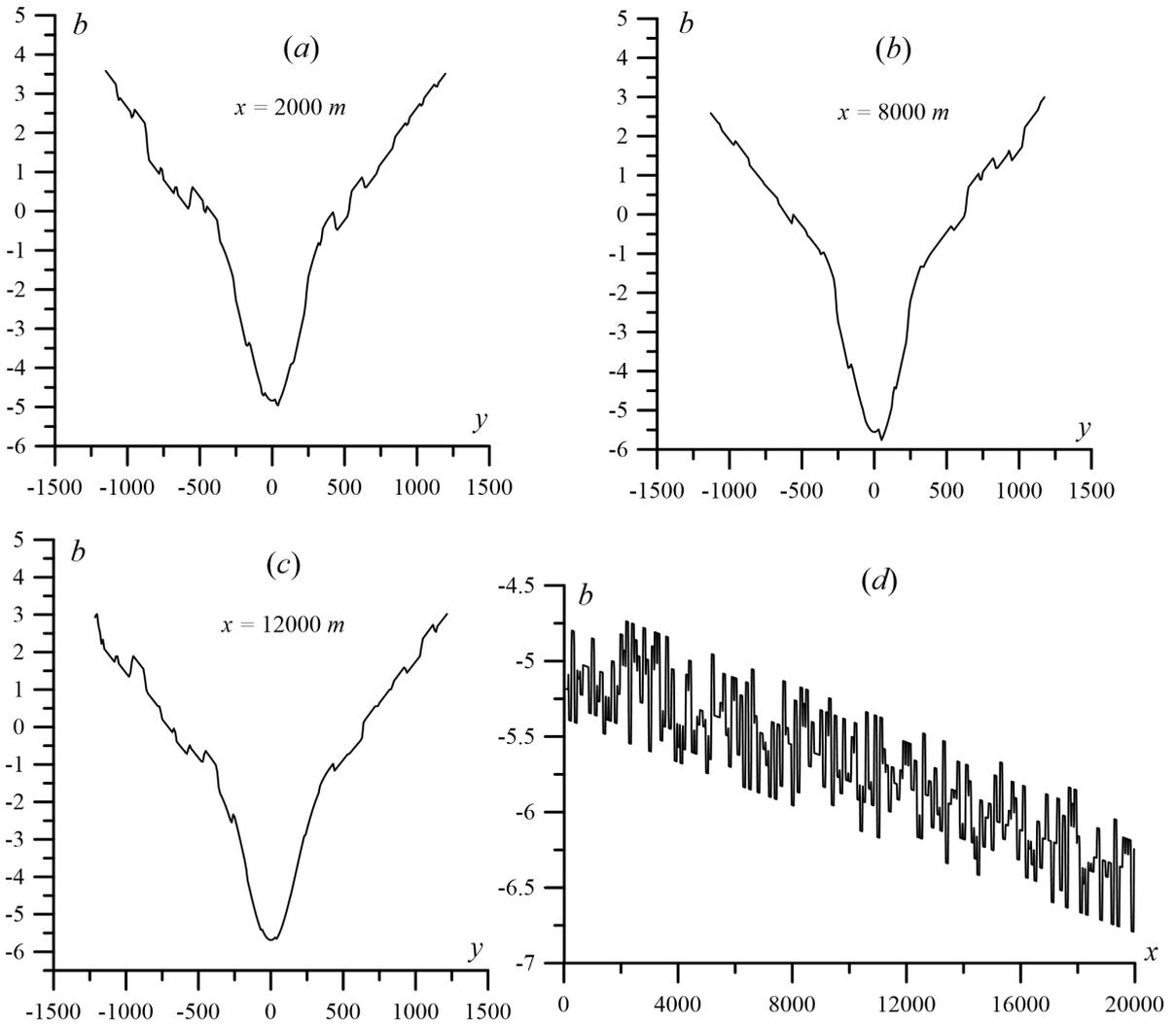}\\
	\end{center}
	\caption{\label{FIG02Dyakonova}The profiles of the perturbed bottom across the channel $ b(y) $ for the three values of the longitudinal coordinate: $x = 2000$~m ({\it a}), $x = 8000$~m ({\it b}), and $x = 12000$~m ({\it c}). The rest of the parameters are $\Delta {b_{\max}} = 0.4$~m, $\ell = 100$~m. The panel ({\it d}) demonstrates the dependence of $b(x)$  for fairway line $y = 0$.}
\end{figure}

\subsection{The problem of determining the Manning coefficient in the Chezy model}

The Chezy model takes into account the connection between the average flow velocity $V_{av}$ and the hydraulic slope $I$, the hydraulic radius $R$, as well as the Chezy coefficient $C$ which describes the frictional resistance:
\begin{equation}\label{eqDyakonovaVav}
V_{av}=C(IR)^{1/2} \,.
\end{equation}

The Pavlovsky formula $C=R^{r}/n_M$  connects the parameter $C$ with the roughness coefficient (or the Manning coefficient) $n_M$ which leads to the Manning formula in the particular case $r = 1/6$:
\begin{equation}\label{eqDyakonovaVav2}
V_{av}= \frac{R^{2/3}I^{1/2}}{n_M} \,.
\end{equation}
The other kind of the Manning's formula for the discharge is
\begin{equation}\label{eq_QManning_Dyakonova}
 Q = \frac{S_n R^{2/3}}{n_M} \sqrt{I} \,,
\end{equation}
where $S_n$ is the area of the wetted watercourse cross-section.

The bottom roughness coefficient for the hydraulic resistance value calculations should be represented in the form \cite{khrapov2013mech,Doncker2009,Hadiani2013,Herget2010}:
\begin{equation}\label{eqDyakonova_nM}
n_{M} = \sum\limits_{k=0}^{K}\,n_{Mk} \,,
\end{equation}
accounting for $K$ of various physical factors which characterize the underlying surface of the bottom and the morphology of the watercourse. These quantities, as a rule, are empirical parameters and there are some problems with their determination \cite{Baryshnikov2010}.
The parameter $n_{M0} $ is defined by the properties of bottom itself on very small scales $\ell \ll H_{av}$ ($H_{av}$ is the average depth along the transverse profile; for typical large rivers it lies within $2 \div 7$~m).
In the case (\ref{eqDyakonova_nM}), we can consider the effect on real bottom roughness caused by small-scale irregular watercourse inhomogeneity ($n_1$), braking caused by vegetation ($n_2$), meandered channel ($n_3$), channel width changes ($n_4$), turbulence ($n_5$), impurity transfer and sediment movement ($n_6$), and etc.
Experimental measurements of the parameter $n_N$ for grass have given the result $n_M = 0.031\div0.035$ \cite{jakubis2000contribution}.

There is a large number of papers focused on the Manning coefficient estimation for various natural objects \cite{AustralianHandbook2009,Motallebian2014,Lang2004,Mohammadi2014,Hatzigiannakis2014,Xia2012}.
 Let us mention results of such research:
\begin{description}
\item[$\bullet$] $n_{M} \approx 0.043$ for the Acheron River, Mitta River, Tambo River; $n_{M} \approx 0.079$ for the Merrimans Creek \cite{Lang2004};
\item [$\bullet$] $n_{M} = 0.02$ and $0.035$ for floodplains of the Sistan Basin \cite{Motallebian2014};
     \item [$\bullet$] $n_{M} = 0.045$ during the wet period and $n_{M} = 0.101$ during the dry period for the Karun River \cite{Mohammadi2014};
     \item [$\bullet$] $n_{M} = 0.027 \div 0.042$ for Promaxonas, $n_{M}$ from 0.134 to 0.942 for Peponia, $n_{M}$ from 0.025 to 0.105 for Amfipoli \cite{Hatzigiannakis2014};
     \item [$\bullet$] $n_{M} = 0.011$ for the Yellow River \cite{Xia2012}.
\end{description}

In the general case, the quantity $n_M$ is inhomogeneous in space. The dependence of $n_M(x, y)$ on the coordinates is caused by the spatial changes in the properties of the underlying surface. This is especially evident in problems of flooding territories or for river streams with extended floodplain.

Let us discuss the factor $n_{M1}$ in (\ref{eqDyakonova_nM}) resulting from the small-scale structure of the bottom heterogeneity \cite{Dyakonova2014}. We have considered the model problem of fluid flow in a channel to study the relationship between the structure of the bottom and the roughness coefficient.

\begin{figure}[th!]
	\begin{center}
		\includegraphics[width=0.85\textwidth]{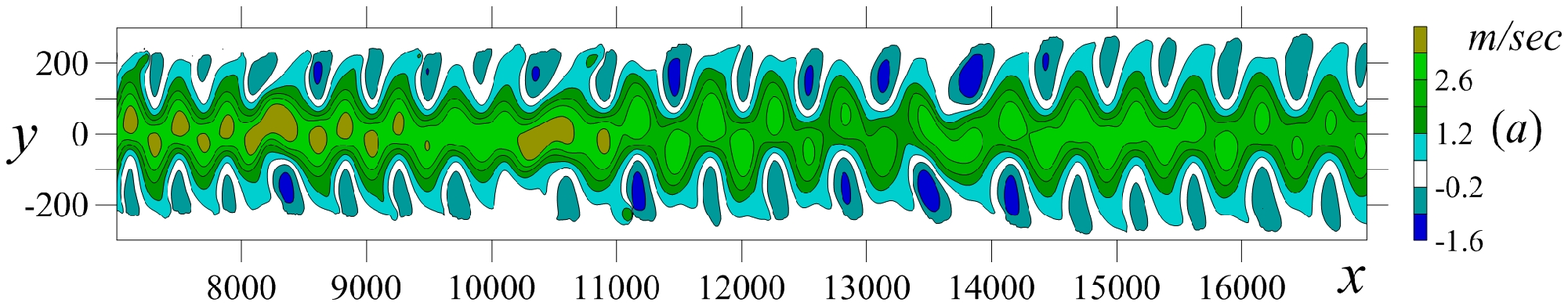}\\
        \includegraphics[width=0.85\textwidth]{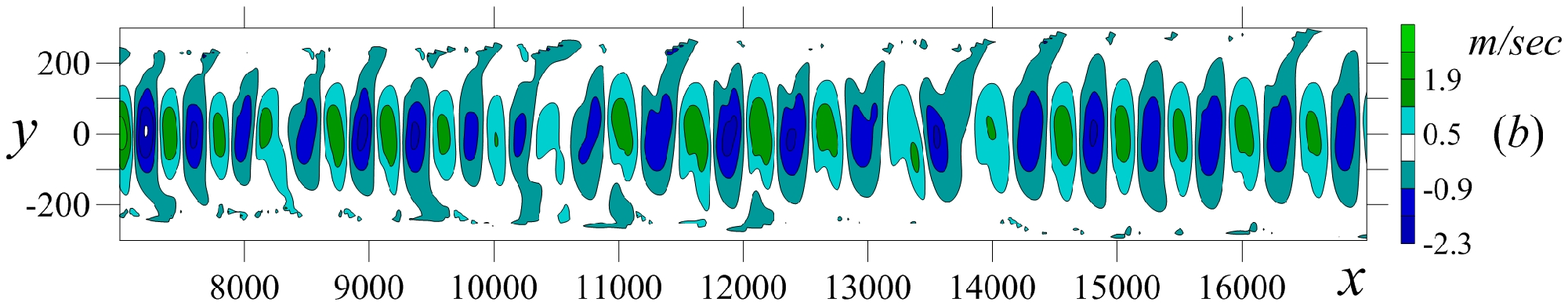}\\
        \includegraphics[width=0.85\textwidth]{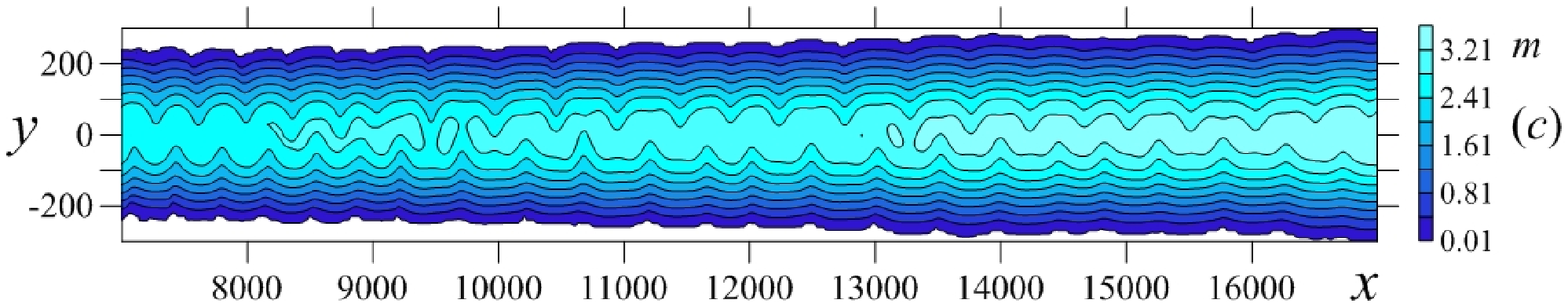}
		\caption{\label{FIG_fragmentcanal_Dyakonova} The spatial distributions of the velocity components $V_x(x, y)$  (\textit{a}), $V_y(x, y)$ (\textit{b}) and depth $H(x, y)$ (\textit{c}) for the channel fragment 7\,km\,$ \le x \le 17$~km with $n_M = 0$. }
	\end{center}
\end{figure}

  The surface of bottom  have been perturbated $b(x, y) = b_0(x, y) + \Delta{b}(x, y)$. The spatial scale $\ell$ and the maximum amplitude $\Delta {b}_{\max}$ are the primary characteristics of perturbation. These perturbations $\Delta{b}(x, y)$ have the form of local hills or valleys depending on the sign of the random variable $\Delta{b}$. To construct the perturbed relief we have chosen the uniformly distributed set of random variables $-\Delta{b}_{\max}  \le \Delta{b} \le \Delta{b}_{\max}$ (Fig.~\ref{FIG02Dyakonova}). We performed a series of numerical experiments on the perturbed reliefs with $\Delta{b}_{\max} = 0.2$~m; 0.4~m; 0.6~m; 0.8~m for the maximum depth of the unperturbed channel $b_{\max} = 5$~m (See Figure~1).

\section{Results and Discussion}

Consider the limiting case $n_M = 0$.
In such a case the water stream is unstable leading to hydrodynamic instability (Figure \ref{FIG_fragmentcanal_Dyakonova}).
Strong transverse motions arise with formation of the vortex structures.
A sequence of non-stationary pulsations is formed along the entire flow (Figure \ref{FIG_fragmentcanal_Dyakonova}).

\begin{figure}[th!]
	\begin{center}
		\includegraphics[width=0.66\textwidth]{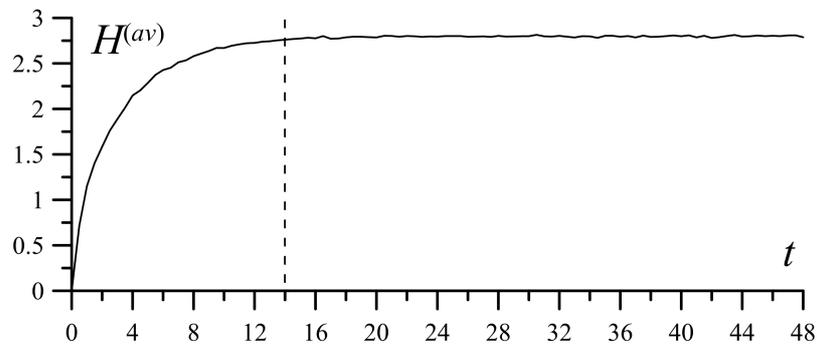}
		\caption{\label{FIG_Ht_nM_Dyakonova} The time dependence of $H^{(av)}$  in the fairway $ y = 0 $ ($[H^{(av)}]=$m, $[t]=$hours)}
	\end{center}
\end{figure}

At $n_M = 0$  the flow becomes substantially two-dimensional due to the appearance of the velocity transverse component $V_y(x, y, t)$  ($\max{V_y} \sim 0.5\max{V_x}$).
In small times and small spatial scales ($\sim L_c$), the flow is nonstationary, the velocity amplitude pulsation of which $\delta {V}$ has the order of the flow velocity $V_{\max}$.
These pulsations ensure the flow braking due to the internal numerical viscosity at small scales ($\sim \Delta {x}$), and as a result, a quasi-stationary state is established.
The velocity profiles averaged over the $x$-coordinate for the cross-section of the channel can be determined as follows:
\begin{equation}\label{eq_Vxav_Dyakonova}
 V_x^{(av)}(y) = \frac{1}{x_{\max}^{(av)}-x_{\min}^{(av)}} \int\limits_{x_{\min}^{(av)}}^{x_{\max}^{(av)}}  V_x(x,y)\,dx \,,
\end{equation}
where $x_{\max}^{(av)} = 20000$~m and $x_{\min}^{(av)} = 50$~m in our models.
The average water depth profile is defined similarly:
\begin{equation}\label{eq_Hav_Dyakonova}
H^{(av)}(y) = \frac{1}{x_{\max}^{(av)}-x_{\min}^{(av)}} \int\limits_{x_{\min}^{(av)}}^{x_{\max}^{(av)}}  H(x,y)\,dx \,.
\end{equation}
After $t>14$~hrs, the profiles $H^{(av)}(y)$, $V_x^{(av)} (y)$ become almost stationary (Figure~\ref{FIG_Ht_nM_Dyakonova}).

\begin{figure}[th!]
	\begin{center}
		\includegraphics[width=0.9\textwidth]{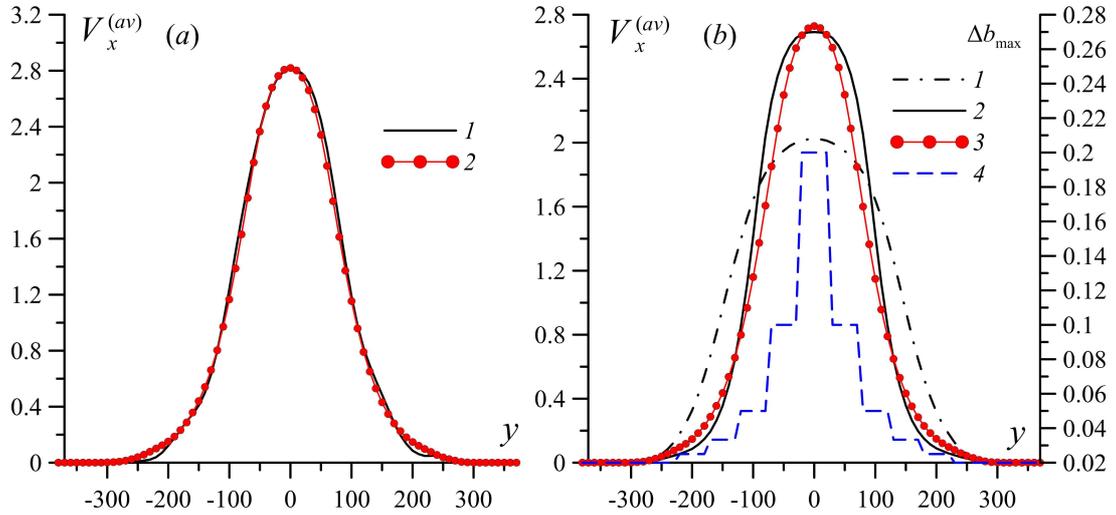}\\
		\caption{\label{FIG_Prof_Dyakonova} The result of concordance of the velocity profiles $V_x^{(av)} (y)$ for the model with perturbed bottom not accounting for bottom friction ($n_M = 0$) (curve \textit{1}) and for the flow on the smooth bottom relief $b_0(x, y)$ with $n_M> 0$ (curve \textit{2}) (\textit{b}). The functions $V_x^{(av)} (y)$ (left axis): the curve {\textit 1} is a model with a constant distribution of the value $\Delta {b_{\max}} (y) = 0.2$~m in the section of the channel; the curve {\textit 2} corresponds to a model with a non-uniform profile $\Delta {b_{\max}} (y)$ (See curve {\textit 4}); the curve {\textit 3} has the best agreement in the case $n_M = 0.008$.
Curve \textit{4} is the distribution of $\Delta {b_{\max}} (y)$ in the riverbed of the channel (right axis) (\textit{b}).}
	\end{center}
\end{figure}

Random inhomogeneities with different amplitude and spatial scale resist to the water flow, and as a result, its average velocity $V_{av}$ decreases. Thus, the transverse velocity profile $V_x(y)$  depends on the perturbed bottom parameters $\Delta{b}(x,y)$. We have conducted calculations with the Manning coefficient $n_M = 0$ for the perturbed bottom. Then we have performed a fitting of the function $V_x^{(av)} (y)$ with the velocity profiles obtained on the smooth (unperturbed) relief $b_0(x, y)$ by varying the parameter $n_M$. Figure~\ref{FIG_Prof_Dyakonova}~{\it  a} shows the result of such  procedure for perturbations of the bottom  with parameters $\ell = 100$~m and $\Delta {b}_{\max} = 0.4$~m.
 Such velocity profiles on the perturbed bottom relief can be reproduced for flows on a smooth bottom with good accuracy for certain values of the Manning coefficient $n_M$. Current approach allows to connect the value of $n_M$ with the heterogeneity parameters of the bottom within the chosen model $\Delta{b}(x, y)$.

For small values of $ \ell\!\!\!\rightarrow\!\!\!\Delta {x} $ the profiles $V_x^{(av)}(y)$ for the models with perturbated bottom  (with $ n_M=0$) and with smoothed bottom (with $n_M\ne0$) are hardly consistent. Figure \ref{FIG_Prof_Dyakonova}~{\it b} demonstrates stronger differences between the solutions for the most small-scale perturbations of the bottom with $\ell=\Delta{x}$, which are limitative in our numerical model of shallow water.

For this case, coincidence can be obtained either for speed along the fairway line, or for the value $\displaystyle \int\limits_{y_{\min}}^{y_{\max}} V_x^{(av)}(y)\,dy$. These methods give different estimates of $n_M$. Therefore, we have also considered the inhomogeneous form of $ n_M(y) $ with a bell-shaped profile (see curve \textit{4} in Figure~\ref{FIG_Prof_Dyakonova}~{\it b}). The latter approach allows the velocity profiles to be in agreement with each other in two different models that indicates the relation between $n_M$ and $H$.

\begin{figure}[th!]
	\begin{center}
		\includegraphics[width=0.55\textwidth]{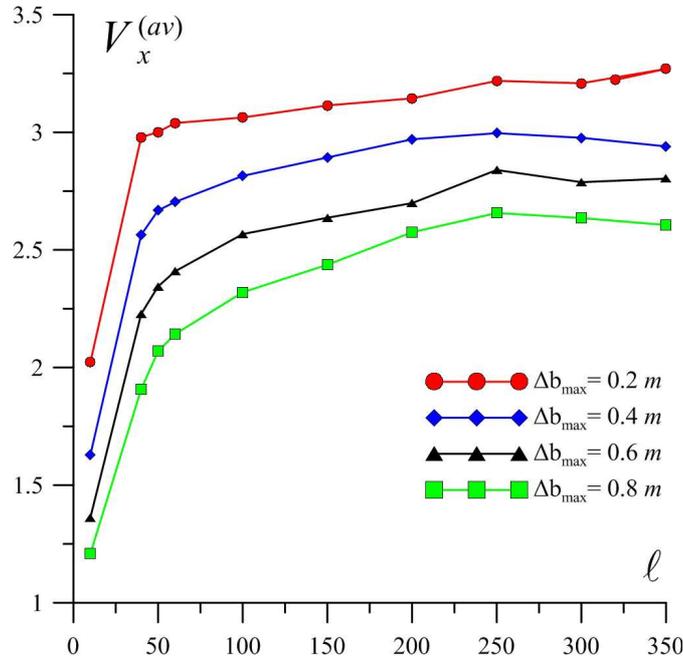}\\
		\caption{\label{FIG_VXprofMidMax_4Dyakonova} The dependence of the maximum velocity at the fairway line $ V_x^{(av)} (y = 0) $~(m/sec) on the inhomogeneity spatial scale  $ \ell $~(m)}
	\end{center}
\end{figure}

More than hundred computational experiments have been conducted for various $\ell$, $\Delta {b}_{\max}$, $n_M$. The latter has allowed establishing a link between the heterogeneity parameters of the bottom and the Manning parameter.
The maximum values of the longitudinal velocity profiles $ V_x^{(av)} (y) $ decrease with increasing of $ n_M $ in the model on smooth (undisturbed) bottom relief, which almost always allows to choose the Manning coefficient for corresponding pertrubated bottom (See Figure~\ref{FIG_VXprofMidMax_4Dyakonova}).

\begin{figure}[h]
	\begin{center}
		\includegraphics[width=0.7\textwidth]{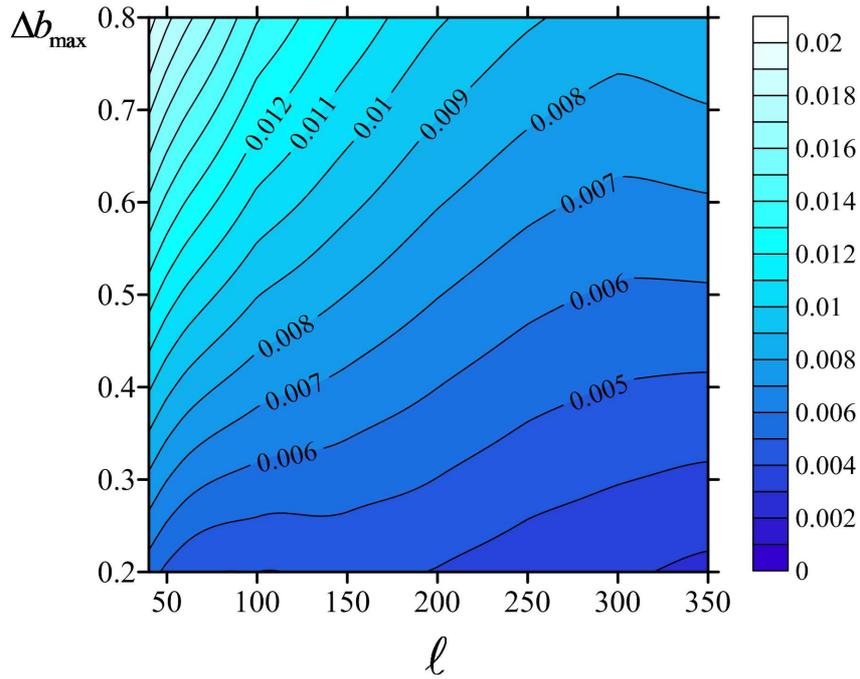}\\
		\caption{\label{FIG_counterNm_Dyakonova} The distribution of the roughness coefficient $n_M$ in the parameter plane $\ell$ (m) and~$\Delta {b}_{\max}$~(m).}
	\end{center}
\end{figure}

Increase of the  maximum amplitude of perturbation of bottom $\Delta {b}_{\max}$  leading to the monotonically increase of the corresponding value of $n_M$. The water flow moves more slowly with increasing of $ \Delta {b}_{\max} $ parameter. The dependence of the maximum flow velocity $ V_x^{(\ max)} $ on scale $ \ell $ has more complicated character (See Figure~\ref{FIG_VXprofMidMax_4Dyakonova}).
The braking of flow is stronger for small-scale perturbations of the bottom which have  an order of magnitude approximately $\ell <50 $~m.
Our estimates of the Manning coefficient for various values of $\Delta {b}_{\max}$ and $\ell$ are presented in the Figure~\ref{FIG_counterNm_Dyakonova}.
For the range of parameters $\ell = 50 \div 350$\,m and $\Delta {b}_{\max} = 0.2 \div 0.8$\,m the Manning coefficient does not exceed 0.02.

\section{Conclusion} %(150-200 слов).
The numerical simulation of hydrodynamic currents of surface water on a realistic terrain relief usually containing a wide range of scales of bottom inhomogenieties has shown that these inhomogeneities exert an additional resistance to the flow.
This factor should be taken into account when choosing the value of the Manning coefficient~$n_M$.
The estimates of $n_M$ accounting for the inhomogeneity parameters of the bottom relief in the channel have been obtained from a series of numerical experiments in the shallow water model.
It should be mentioned that these results largely depend on the adopted model of the surface perturbation of bottom relief which requires additional research.

The results obtained within the traditional shallow water model with a constant Manning coefficient $(n_M = {\textrm{const}})$ are in agreement with the results of our direct numerical experiments on the inhomogeneous bottom relief with scales $\ell \geq 50 $~m. The $ n_M $ dependence on the coordinates should be accounted for better coincidence in the case of small-scale inhomogeneities ($\ell <50 $~m).

\section*{Acknowledgments}
 The authors are grateful to A.Yu. Klikunova, S.S. Khrapov, and A.V. Pisarev for help in the work and useful discussions.
  A. Khoperskov is thankful to the Ministry of Education and Science of the Russian Federation (government task No.2.852.2017/4.6).
  T. Dyakonova is thankful RFBR and Volgograd Region Administration (grant No.~15-45-02655), grant of young researchers of VolSU in 2017 also.
 The research is carried out using the equipment of the shared research facilities of HPC computing resources at Lomonosov Moscow State University.

\section*{References}

\providecommand{\newblock}{}


\begin{thebibliography}{10}
\expandafter\ifx\csname url\endcsname\relax
  \def\url#1{{\tt #1}}\fi
\expandafter\ifx\csname urlprefix\endcsname\relax\def\urlprefix{URL }\fi
\providecommand{\eprint}[2][]{\url{#2}}

\bibitem{shokina2015numerical}
Shokina N and Aizinger V 2015 {\em Environmental Earth Sciences\/} {\bf 74}
  7387--7405

\bibitem{agafonnikova2017computer}
Agafonnikova E~O, Klikunova A~Y and Khoperskov A~V 2017 {\em Bulletin of the
  South Ural State University, Series: Mathematical Modelling, Programming and
  Computer Software.\/} {\bf 10} 148--155

\bibitem{shokin2016combined}
Shokin Y~I, Rychkov A~D, Khakimzyanov G~S and Chubarov L~B 2016 {\em Russian
  Journal of Numerical Analysis and Mathematical Modelling\/} {\bf 31} 217--227

\bibitem{Cui2013}
Cui X and Gray J 2013 {\em Journal of Fluid Mechanics\/} {\bf 720} 314--337

\bibitem{Majda2003}
Majda A 2003 {\em Introduction to PDEs and Waves for the Atmosphere and
  Ocean\/} vol~9 (American Mathematical Soc.)

\bibitem{Luo2017}
Luo Z, Wu Q and Zhang L 2017 {\em Journal of Physics: Conference Series\/} vol
  916 (IOP Publishing) p 012042

\bibitem{kulagin2016physical}
Kulagin V, Moskvichev V, Makhutov N, Markovich D and Shokin Y~I 2016 {\em
  Herald of the Russian Academy of Sciences\/} {\bf 86} 454--465

\bibitem{Baryshnikov2010}
Baryshnikov S~O and Pagin A~O 2010 {\em Vestnik gosudarstvennogo universiteta
  morskogo i rechnogo flota imeni admirala S.O. Makarova\/}  {\bf 2 (6)} 90--93

\bibitem{Dyakonova2014}
Dyakonova T~A, Pisarev A~V, Khoperskov A~V and Khrapov S~S 2014 {\em Science
  Journal of Volgograd State University. Mathematics. Physics\/}  {\bf 20} 35--44

\bibitem{Pisarev2013}
Pisarev A~V, Khrapov S~S, Agafonnikova E~O and Khoperskov A~V 2013 {\em
  Bulletin of Udmurt University. Mathematics, Mechanics\/} {\bf 1} 114--130

\bibitem{Jong-Seok2017}
Jong-Seok L and Pierre Y~J 2017 {\em Journal of flood engineering\/}  {\bf 8 (2)} 55--75

\bibitem{habibi2014experimental}
Habibi M, Namaee M~R and Saneie M 2014 {\em KSCE Journal of Civil
  Engineering\/} {\bf 18} 1176--1184

\bibitem{Doncker2009}
De~Doncker L, Troch P, Verhoeven R, Bal K, Meire P and Quintelier J 2009 {\em
  Environmental fluid mechanics\/} {\bf 9} 549--567

\bibitem{Hadiani2013}
Hadiani M, Asl S~J, Banafsheh M~R and Dinpajouh Y 2013 {\em World Applied
  Sciences Journal\/} {\bf 22} 307--312

\bibitem{Herget2010}
Herget J and Meurs H 2010 {\em Global and Planetary Change\/} {\bf 70} 108--116

\bibitem{Mohammadi2014}
Mohammadi S and Kashefipour S~M 2014 {\em Water resources\/} {\bf 41} 412--420

\bibitem{Hatzigiannakis2016}
Hatzigiannakis E, Pantelakis D, Hatzispiroglou I, Arampatzis G, Ilias A and
  Panagopoulos A 2016 {\em Environmental Processes\/} {\bf 3} 263--275

\bibitem{Hatzigiannakis2014}
Hatzigiannakis E, Pantelakis D, Hatzispiroglou I, Arampatzis G, Ilias A and
  Panagopoulos A 2014   888--895

\bibitem{Lyra2010}
Lyra G~B, Cecilio R~A, Zanetti S~S and Lyra G~B 2010 {\em Revista Brasileira de
  Engenharia Agricola e Ambiental-Agriambi\/} {\bf 14}

\bibitem{khrapov2013mech}
Khrapov S~S, Pisarev A~V, Kobelev I~A, Zhumaliev A~G, Agafonnikova E~O, Losev
  A~G and Khoperskov A~V {\em Advances in Mechanical Engineering\/} {\bf 5}
  787016

\bibitem{nistoran2017one}
Nistoran D~G, Ionescu C, Patru G, Armas I and Omrani S~G 2017 {\em Energy
  Procedia\/} {\bf 112} 67--74

\bibitem{Voronin2017}
Voronin A, Isaeva I, Khoperskov A and Grebenjuk S 2017 {\em Conference on
  Creativity in Intelligent Technologies and Data Science\/} (Springer) pp
  419--429

\bibitem{Dyakonova2016boundary}
D'yakonova T, Khrapov S and Khoperskov A 2016 {\em Vestnik Udmurtskogo
  Universiteta: Matematika, Mekhanika, Komp'yuternye Nauki\/} {\bf 26} 401--417

\bibitem{Khrapov2011}
Khrapov S~S, Khoperskov A~V, Kuz'min N~M, Pisarev A~V and Kobelev I~A 2011 {\em
  Vychisl. Metody Programm\/} {\bf 12} 282--297

\bibitem{Dyakonova2016}
Dyakonova T, Khoperskov A and Khrapov S 2016 {\em Communications in Computer
  and Information Science\/} {\bf 687} 132--145

\bibitem{jakubis2000contribution}
Jakubis M 2000 {\em Journal of the International Commission on Irrigation and
  Drainage\/} {\bf 49} 41--54

\bibitem{AustralianHandbook2009}
Lang S, Ladson A, Anderson B and Rutherford I 2004 {\em Stream roughness, Four
  case studies from Victoria, AJWR\/} {\bf 28}

\bibitem{Motallebian2014}
Motallebian M and Hassanpour F 2014 {\em Journal of Civil Engineering and
  Urbanism\/}  {\bf 4 (5)} 540--545

\bibitem{Lang2004}
Lang S, Ladson T, Anderson B and Rutherford I 2004 {\em National Rivers
  Consortium: Canberra\/}

\bibitem{Xia2012}
Xia J, Lin B, Falconer R~A and Wang Y 2012 {\em Proceedings of the Institution
  of Civil Engineers\/} {\bf 165} 377

\end{thebibliography}
\end{document}